\documentclass[a4paper]{article}

\usepackage{INTERSPEECH2019}

\title{Multi-Task Learning with High-Order Statistics for X-vector based Text-Independent Speaker Verification}
\name{Lanhua You, Wu Guo, Lirong Dai, Jun Du}
\address{
National Engineering Laboratory for Speech and Language Information Processing\\
University of Science and Technology of China, Hefei, China
}
\email{lhyou@mail.ustc.edu.cn, \{guowu,lrdai,jundu\}@ustc.edu.cn}

\begin{document}

\maketitle
\begin{abstract}
The x-vector based deep neural network (DNN) embedding systems have demonstrated effectiveness for text-independent speaker verification. This paper presents a multi-task learning architecture for training the speaker embedding DNN with the primary task of classifying the target speakers, and the auxiliary task of reconstructing the first- and higher-order statistics of the original input utterance. The proposed training strategy aggregates both the supervised and unsupervised learning into one framework to make the speaker embeddings more discriminative and robust. Experiments are carried out using the NIST SRE16 evaluation dataset and the VOiCES dataset. The results demonstrate that our proposed method outperforms the original x-vector approach with very low additional complexity added.
\end{abstract}
\noindent\textbf{Index Terms}: Speaker verification, High-order statistics, X-vector, Multi-task learning, Unsupervised learning

\section{Introduction}

Speaker verification (SV) is the task of verifying a person's claimed identity from speech signals. Converting speaker-specific characteristics from variable length utterances into fixed length vectors is a key component of these SV systems. Over the years, low-dimensional embedding based systems have been widely used for text-independent speaker verification. Usually, the embeddings are represented by i-vectors \cite{dehak2011front}. Combined with the probabilistic linear discriminant analysis (PLDA) \cite{kenny2010bayesian} backend, the i-vector/PLDA framework has become the dominant approach for the last decade.

With the great success of deep neural networks (DNNs) in machine learning fields, many novel DNN frameworks have been proposed to extract the speaker embeddings which can achieve comparable or even better performance compared with the traditional i-vector/PLDA system \cite{variani2014deep, heigold2016end, snyder2017deep, bhattacharya2017deep, zhang2017end}. Typically, the utterance-level optimized speaker embeddings, such as x-vectors \cite{snyder2017deep, snyder2018x}, are used to replace i-vectors as the front-end and a trainable backend (such as PLDA) is employed for further modeling.

Most speaker embedding extraction only utilizes speaker labels and do not consider other information in the model training. Since the phonetic information is the predominant component in speech signals, some researchers have tried to add this information in model training using multi-task learning (MTL) frameworks \cite{liu2018speaker, chen2015multi, tang2017collaborative}. However, most utterances do not have phonetic labels at hand and an additional automatic speech recognition (ASR) system is always required. This requirement will result in longer training times. Furthermore, the recognition errors in some utterances with low signal-to-noise ratios (SNR) may limit the use of the phonetic labels for speaker verification.

In this paper, we propose a novel multi-task learning strategy based on the x-vector architecture with very low complexity added. We first compute the first- and higher-order statistics (including the mean, standard deviation, skewness and kurtosis) of an input utterance, and these four statistics are then used as the reconstruction targets of the auxiliary task. The high-order statistics (HOS) contain statistical characteristics of the input signal and are more easily obtained compared with other auxiliary information. With the MTL, the deep embedding network is trained not only to classify the target speakers but also to reconstruct both the low- and high-order statistics from input features. The resulting speaker embeddings can benefit from both the supervised and unsupervised learning. We evaluate our experiments using the NIST SRE16 evaluation dataset \cite{sadjadi20172016} and the VOiCES dataset \cite{nandwana2019voices}. The experimental results show that our proposed method achieves better performance compared with the original x-vector approach and requires very little extra computation load.

The remainder of this paper is organized as follows. Section 2 gives an introduction to our x-vector baseline. Section 3 presents the high-order statistics, as well as the proposed MTL strategy. Section 4 presents the experimental setup and the results of this study. In Section 5, we summarize our work and discuss future work.

\section{Baseline network architecture}

The network architecture of our x-vector baseline system is the same as that described in \cite{snyder2017deep}. As depicted in Figure 1, the first five TDNN (or 1-dimensional dilated CNN) layers ${l_1}$ to ${l_5}$ are stacked for extracting the frame-level features. More specifically, the TDNN layers with dilation rates of 2 and 3 are used for the second and third layers respectively, while the others retain the dilation rate of 1. The kernel sizes of the five layers are 5, 3, 3, 1 and 1, respectively.

The final frame-level output vectors of the whole variable-length utterance are aggregated into a fixed segment-level vector through the statistics pooling layer. The mean and standard deviation are calculated and then concatenated together for the statistics pooling. Two additional fully connected layers ${l_6}$ and ${l_7}$ are added to obtain a low-dimensional utterance-level representation that is finally passed into a softmax output layer.

The deep embedding network is trained to predict the correct speaker labels with cross entropy (CE) loss. Once the DNN is trained, we remove the softmax layer and the last fully connected layer ${l_7}$, and the output of the linear affine layer directly on top of the statistics pooling is extracted as the speaker embedding.
\begin{figure}[t]
  \centering
  \includegraphics[width=0.7\linewidth, height=0.7\linewidth]{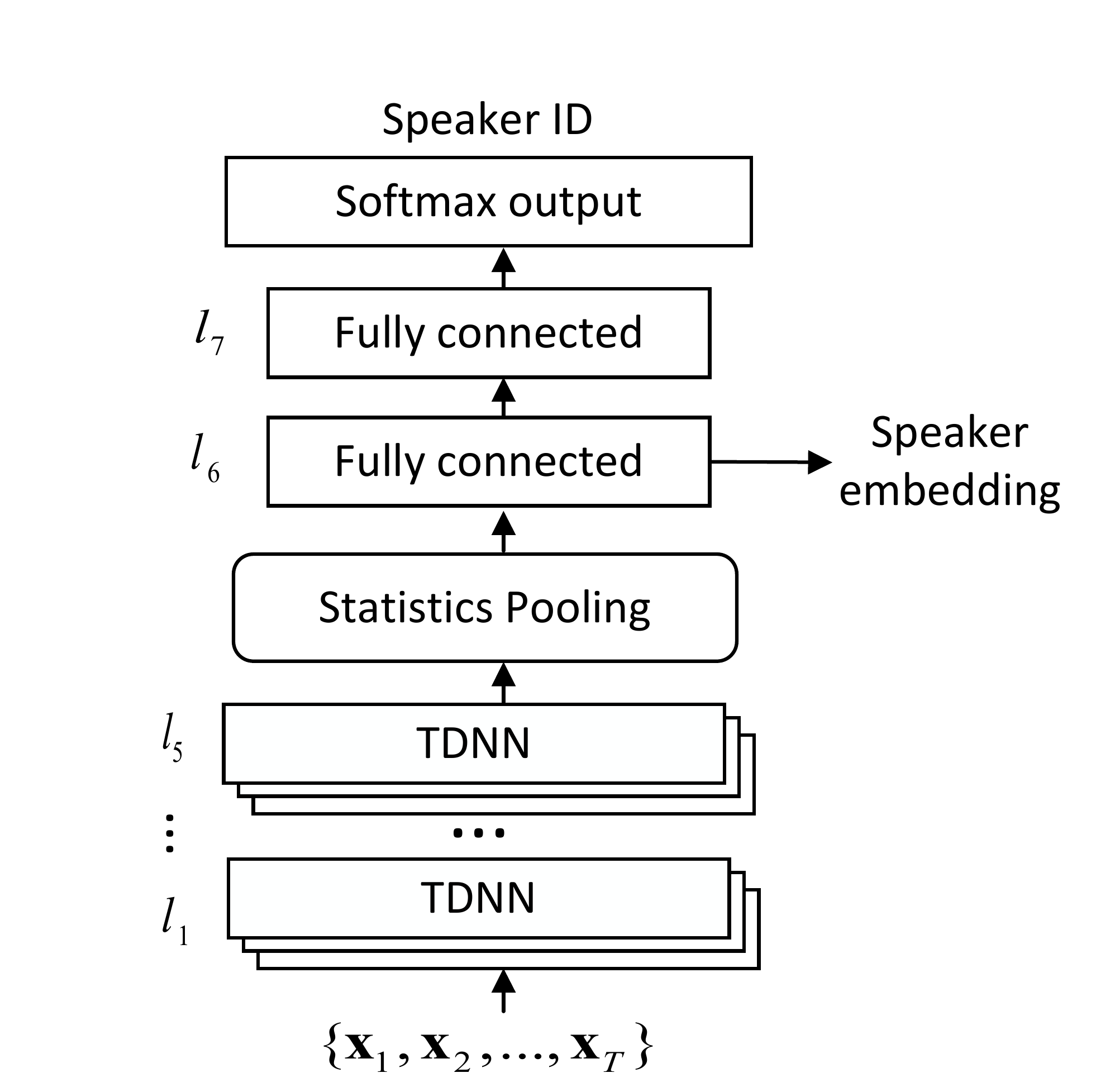}
  \caption{Network architecture of the x-vector baseline.}
  \label{fig:x-vector baseline}
\end{figure}

\section{Multi-task learning with high-order statistics}

\subsection{High-order statistics}
Higher-order statistics can be used in estimation of the shape of unimodal distributions and have been applied to many tasks \cite{xu2016blind, richiardi2008evaluation, molla2004effectiveness}. In the original x-vector system, low-order statistics, such as the mean and standard deviation, are calculated to perform the statistics pooling and have demonstrated their effectiveness for extracting speaker-specific embeddings \cite{snyder2017deep, Okabe2018}. Here, these statistics of input features are computed as the fixed-length input representations. Given an input utterance, ${\bf{X}} = [{{\bf{x}}_1},{{\bf{x}}_2},...,{{\bf{x}}_T}]$, the different order statistics can be calculated as follows
\begin{equation}
 {\bm{\mu }} = \frac{1}{T}\sum\limits_{t = 1}^T {{{\bf{x}}_t}}
\end{equation}
\begin{equation}
{\bm{\sigma }} = \sqrt {\frac{1}{T}\sum\limits_{t = 1}^T {({{\bf{x}}_t} - {\bm{\mu }}} {)^2}}
\end{equation}
\begin{equation}
{\bf{s}} = \frac{1}{T}\sum\limits_{t = 1}^T {(\frac{{{{\bf{x}}_t} - {\bm{\mu }}}}{{\bm{\sigma }}}} {)^3}
\end{equation}
\begin{equation}
{\bf{k}} = \frac{1}{T}\sum\limits_{t = 1}^T {(\frac{{{{\bf{x}}_t} - {\bm{\mu }}}}{{\bm{\sigma }}}} {)^4}
\end{equation}
where $T$ is the number of frames in the input utterance. ${\bm{\mu }}$ and ${\bm{\sigma }}$ are the mean and standard deviation vectors, respectively. In addition to the first- and second-order statistics, higher-order statistics including skewness ${\bf{s}}$ and kurtosis ${\bf{k}}$, are also used to describe the statistical characteristics of the input utterance. The skewness measures the asymmetry of a distribution with respect to its mode while the kurtosis measures the ``tailedness" of the data distribution. Both of these values enrich the statistical information from input features. Finally, we concatenate these four statistics into a fixed-dimensional vector, and we call this the HOS vector in this paper.
\begin{equation}
{\bf{z}} = [{\bm{\mu }},{\bm{\sigma }},{\bf{s}},{\bf{k}}]
\end{equation}

\subsection{Multi-task learning strategy}
In the original x-vector architecture, only speaker labels are considered for the DNN training. We use a MTL strategy to incorporate the abovementioned HOS vector into the x-vector DNN training, where the classification of the speaker label is still the primary task, and the reconstruction of the HOS vector $\bf{z}$ is the auxiliary task. As depicted in Figure 2, the speaker label is a supervised label and the HOS vector is an unsupervised label; therefore, the proposed MTL training strategy aggregates both the supervised and unsupervised learning into one framework.

\begin{figure}[t]
  \centering
  \includegraphics[width=\linewidth]{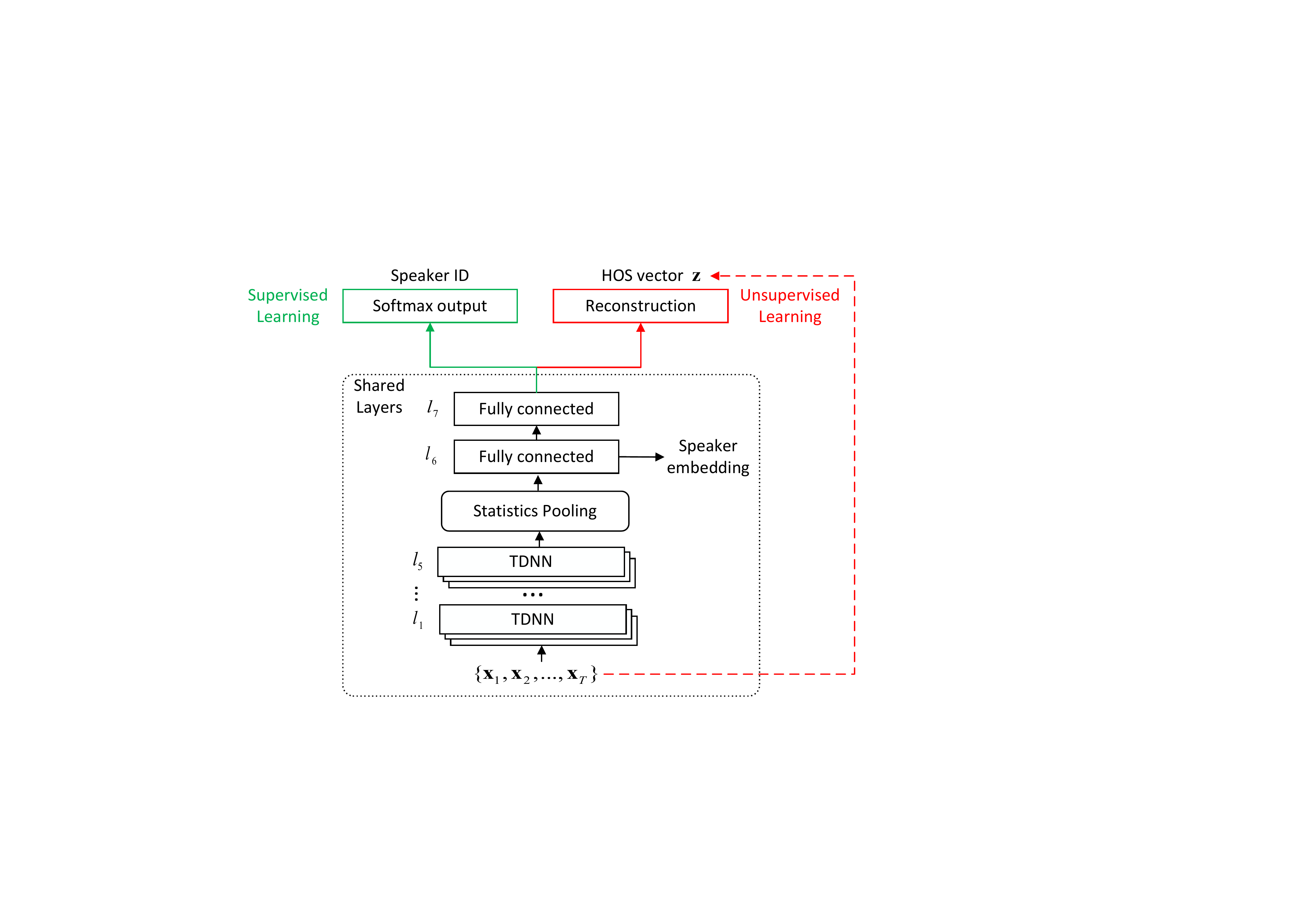}
  \caption{The proposed multi-task learning architecture.}
  \label{fig:multi task learning}
\end{figure}

In MTL, the network is trained to perform both the primary classification task and the auxiliary reconstruction task. Except for the task-specific output layers, all the layers are shared between both tasks. For unsupervised learning, we add a linear layer on top of the last shared layer $l_7$ to obtain the reconstructed vector $\bf{h}$. The reconstruction task aims to minimize the mean square error (MSE) loss between its output vector and $\bf{z}$, which can be formulated as
\begin{equation}
MSE = \frac{1}{N}\sum\limits_{n = 1}^N {{{\left\| {{{\bf{h}}^{(n)}} - {{\bf{z}}^{(n)}}} \right\|}^2}}
\end{equation}
where ${N}$ is the total number of training samples and ${{\bf{h}}^{(n)}}$ is the reconstruction of statistical representation ${{\bf{z}}^{(n)}}$ for the $n{\rm{-th}}$ input utterance. Combined with the original CE loss, the final loss function can be written as
\begin{equation}
Loss = \alpha MSE + (1 - \alpha )CE
\end{equation}
where $\alpha$ is the task weight.

After the multi-task training, the extracted speaker embedding will contain both the discriminative and unsupervised speaker information. From a model training perspective, the auxiliary task enhances the model generalization ability by introducing regularization into the shared layers. Since the vector $\bf{z}$ is quite low-dimensional (4 $ \times $ feature dimension), adding the auxiliary task only slightly increases the number of parameters in the output layer and requires very low extra computational cost. Moreover, the extra overhead can be neglected, since the top layers are removed when extracting speaker embeddings.

\begin{table*}[th]
  \caption{Results of the NIST SRE16. MT-o$n$-${\alpha}$$p$ denotes the multi-task learning system with task weight ${\alpha=p/10}$ and it concatenates the statistics from 1 to $n$ orders to perform the HOS vector ${\bf{z}}$. For example, MT-o$3$-$\alpha$$3$ means ${\bf{z}} = [{\bm{\mu }},{\bm{\sigma }},{\bf{s}}]$ and $\alpha  = 0.3$. Here, we focus on the performance with respect to weight ${\alpha}$.}
  \label{tab:table1}
  \centering
  \setlength{\tabcolsep}{3mm}
  \begin{tabular}{ccccccccccc}
    \toprule
    & \multicolumn{3}{c}{SRE16, Pooled} & \multicolumn{3}{c}{SRE16, Cantonese} & \multicolumn{3}{c}{SRE16, Tagalog}\\
    \cmidrule(r){2-4} \cmidrule(r){5-7} \cmidrule(r){8-10}
    system      & EER\%& minDCF& actDCF& EER\%& minDCF& actDCF& EER\% & minDCF& actDCF\\

    \toprule
    x-vector             & 8.03      & 0.586      & 0.605      & 4.06       & 0.396      & 0.404      & 12.00       & 0.743         & 0.807\\ \hline
    MT-o4-${\alpha}$1    & 8.09      & 0.581      & 0.607      & 4.04       & 0.379      & 0.382      & 12.15       & 0.744         & 0.832\\
    MT-o4-${\alpha}$2    & 7.91      & 0.571      & 0.583      & \bf{3.83}  & 0.390      & 0.398      & 11.99       & \bf{0.723}     & 0.768\\
    MT-o4-${\alpha}$3    & \bf{7.79} & \bf{0.563} & \bf{0.568} & 3.88       & 0.368      & 0.376      & \bf{11.69}  & 0.732         & \bf{0.761}\\
    MT-o4-${\alpha}$4    & 8.03      & 0.564      & 0.577      & 3.90       & \bf{0.364} & \bf{0.375} & 12.18       & 0.727         & 0.779\\ \hline
    MT-o4-${\alpha}$10   & 26.3      & 0.979      & 4.40       & 22.7       & 0.961      & 4.02       & 29.9        & 0.990         & 4.79\\

    \bottomrule

  \end{tabular}
\end{table*}

\begin{table*}[th]
  \caption{Results of the NIST SRE16. Comparison results of proposed multi-task learning systems using different orders of statistics when ${\alpha}=0.3$.}
  \label{tab:table2}
  \centering
  \setlength{\tabcolsep}{3mm}
  \begin{tabular}{ccccccccccc}
    \toprule
    & \multicolumn{3}{c}{SRE16, Pooled} & \multicolumn{3}{c}{SRE16, Cantonese} & \multicolumn{3}{c}{SRE16, Tagalog}\\
    \cmidrule(r){2-4} \cmidrule(r){5-7} \cmidrule(r){8-10}
    system      & EER\%& minDCF& actDCF& EER\%& minDCF& actDCF& EER\% & minDCF& actDCF\\

    \toprule
    MT-o1-${\alpha}$3    & 7.85     & 0.578     & 0.593     & \bf{3.76} & 0.370         & \bf{0.376}    & 11.99         & 0.751         & 0.810\\
    MT-o2-${\alpha}$3    & 8.31     & 0.568     & 0.575     & 4.18      & 0.378         & 0.386         & 12.39         & 0.736         & 0.764\\
    MT-o3-${\alpha}$3    & 8.05     & 0.572     & 0.585     & 3.89      & 0.376         & 0.386         & 12.21         & \bf{0.730}    & 0.783\\
    MT-o4-${\alpha}$3    & \bf{7.79}& \bf{0.563}& \bf{0.568}& 3.88      & \bf{0.368}    & \bf{0.376}    & \bf{11.69}    & 0.732         & \bf{0.761}\\

    \bottomrule

  \end{tabular}
\end{table*}

\section{Experiments and analysis of results}
\subsection{Experimental settings}
All systems are based on the TensorFlow implementation of the x-vector speaker embedding \cite{zeinali2018improve}. We trained the network and extracted x-vectors using TensorFlow $\footnote{https://github.com/hsn-zeinali/x-vector-kaldi-tf/tree/master/local/tf}$. The other procedures (including data processing, feature extraction and PLDA backend) are implemented using Kaldi Toolkit \cite{snyder2017deep, snyder2018x}.
\subsubsection{Training data and evaluation metric}

The experiments are carried out on the NIST SRE16 evaluation dataset, and both the development and evaluation parts of the VOiCES dataset. For the NIST SRE16 dataset, the training data mainly consists of the telephone speech (with a small amount of the microphone speech) from the NIST SRE2004-2010, Mixer 6 and Switchboard datasets. We also use the data augmentation techniques described in \cite{snyder2018x}, and employ the babble, music and noise augmented data to increase the amount and diversity of the existing training data. In summary, there are a total of 183,457 recordings from 7,001 speakers, including approximately 96,000 randomly selected augmented recordings.

The VOiCES dataset for the speaker verification task is described in the ``VOiCES from a Distance Challenge 2019" \cite{nandwana2019voices}. The VOiCES development dataset contains 15,904 segments of noisy and far-field speech from 196 speakers. The evaluation set consists of 11,392 distant recordings from different microphone types and different rooms, both of which could be more challenging than those featured in the development set. We use both the Voxceleb1 and Voxceleb2 datasets \cite{Nagrani2017, Chung2018} as the training set for VOiCES Challenge. Data augmentation techniques (including music, noise and reverberation) are also applied for model training.

The performance is evaluated in terms of equal error rate (EER), the minimal detection cost function (minDCF) and actual detection cost function (actDCF) calculated using the SRE16 and VOiCES official scoring softwares. For the NIST SRE16, the equalized results are used.

\subsubsection{Input fearures}

For the NIST SRE16, our input acoustic features are 23-dimensional MFCC features with a frame-length of 25 ms that are mean-normalized over a sliding window of up to 3 s. An energy-based VAD is used to filter out nonspeech frames. Instead of 23-dimensional MFCCs, we use the 30-dimensional same type of MFCCs for VOiCES.

\subsubsection{Model configuration}

In all x-vector based systems, for both the SRE16 and VOiCES, the number of hidden nodes for the first four frame-level layers is 512, while that number is 1536 for the last frame-level layer. Each of the two fully connected layers $l_6$ and $l_7$ has 512 nodes. All the nonlinear activation functions of hidden layers are ReLU. We use the same type of batch normalization and L2 weight decay as in \cite{zeinali2018improve} to prevent overfitting.

\subsubsection{PLDA backend}

For the NIST SRE16, the DNN embeddings are centered using the unlabeled development data and are projected using LDA, which reduces the dimensionality of x-vectors to 100. For training the PLDA model, all the training data (except the Switchboard data) and their corresponding augmented versions are used. Finally, the PLDA model is adapted to the unlabeled data through the unsupervised adaptation in Kaldi. For VOiCES, no adaptation technique is used. We select the longest 200,000 recordings from the training set to train the backend, and the best LDA dimension is 110.

\subsection{Results and analysis}
\subsubsection{NIST SRE16}

Table 1 presents the performance of MTL systems with different task weights in NIST SRE16. It can be observed that the proposed MTL systems with ${\alpha  = 0.2\sim0.4}$ outperform the x-vector baseline system (i.e., ${\alpha  = 0}$ in multi-task learning). When ${\alpha  = 0.3}$, we can obtain the best performance, which is better than the x-vector baseline in terms of all evaluation metrics. On Cantonese, this strategy provides a 4\% relative improvement in terms of EER and a 7\% improvement in terms of both minDCF and actDCF over the original x-vector. The results demonstrate the effectiveness of the proposed MTL strategy. Note that we still can obtain an EER of 22.7\% on Cantonese even when ${\alpha  = 1.0}$. This result shows that embeddings still contain speaker-discriminative information using only our unsupervised learning and shows the importance of unsupervised information.

In Table 2, we investigate the effect of different orders of statistics in multi-task learning where ${\alpha  = 0.3}$. On average, adding both the skewness and kurtosis can achieve better performance. This result demonstrates that speaker embeddings can benefit from the higher-order statistics.

\subsubsection{VOiCES}
For VOiCES, we directly use $\alpha  = 0.3$, which is tuned in the NIST SRE16. The results of VOiCES are reported in Table 3 and Table 4.
\begin{table}[th]
  \caption{Results on the VOiCES development where ${\alpha}=0.3$.}
  \label{tab:table3}
  \centering
  \setlength{\tabcolsep}{3mm}
  \begin{tabular}{cccc}
    \toprule
    system      & EER\%& minDCF& actDCF\\

    \toprule

    x-vector             & 3.36          & 0.387          & 0.515\\ \hline
    MT-o1-${\alpha}$3    & 3.22          & 0.369          & 0.495\\
    MT-o2-${\alpha}$3    & \bf{2.97}     & \bf{0.335}     & \bf{0.409}\\
    MT-o3-${\alpha}$3    & 3.13          & 0.369          & 0.443\\
    MT-o4-${\alpha}$3    & 3.37          & 0.373          & 0.477\\

    \bottomrule

  \end{tabular}
\end{table}
\begin{table}[th]
  \caption{Results on the VOiCES evaluation where ${\alpha}=0.3$.}
  \label{tab:table4}
  \centering
  \setlength{\tabcolsep}{3mm}
  \begin{tabular}{cccc}
    \toprule
    system      & EER\%& minDCF& actDCF\\

    \toprule

    x-vector             & 8.25      & 0.628      & 0.779\\ \hline
    MT-o1-${\alpha}$3    & 7.90      & 0.595      & 0.709\\
    MT-o2-${\alpha}$3    & 8.02      & 0.590      & 0.702\\
    MT-o3-${\alpha}$3    & 7.89      & 0.603      & 0.687\\
    MT-o4-${\alpha}$3    & \bf{7.66} & \bf{0.572} & \bf{0.639}\\

    \bottomrule

  \end{tabular}
\end{table}

We can observe a more obvious improvement provided by our proposed MTL strategy on both development and evaluation. It is interesting to see that the MTL system using only the mean and standard deviation can achieve the best performance for the development set and can improve on the x-vector baseline system by 12\% in EER, 13\% in minDCF and 21\% in actDCF. For the more challenging evaluation set, using all four orders statistics in multi-task learning can provide the largest improvement and outperform the original x-vector by 7\% in EER, 9\% in minDCF and 18\% in terms of actDCF. These results make clear that our MTL strategy makes the speaker embedding more discriminative and robust by joint learning speaker classification and reconstruction of the high-order statistics.

\subsubsection{Speed}

Here, we compare the training speed of the original network with that of the multi-task training network. The networks are trained on Nvidia GeForce GTX 1080Ti GPU. Table 5 presents the time required for each iteration when the two networks are trained using the Voxceleb datasets. It can be observed that MT-o4-$\alpha$3, in which ${\bf{z}} = [{\bm{\mu }},{\bm{\sigma }},{\bf{s}},{\bf{k}}]$ ( $4 \times 30 = 120$ dimensional), is only 4\% slower than the original x-vector. For the other case when not all statistics are used or the dimensionality of input features is smaller than 30, our MTL systems could be faster. This result demonstrates that our multi-task learning strategy improves the performance with very low additional complexity added.
\begin{table}[th]
  \caption{Training speed comparison.}
  \label{tab:table5}
  \centering
  \setlength{\tabcolsep}{3mm}
  \begin{tabular}{cc}
    \toprule
    system              &       Time(min)/Iter\\
    \toprule
    x-vector            &          10.052\\
    MT-o4-${\alpha}$3   &          10.476\\

    \bottomrule

  \end{tabular}
\end{table}
\section{Conclusions}

In this study, we propose a novel MTL strategy for x-vector based architecture. The network is trained not only to classify the target speaker but also to reconstruct the HOS vector of input utterance. The experimental results demonstrate the effectiveness of our proposed strategy compared with the original x-vector architecture. The speaker embeddings will benefit from the auxiliary HOS information and can be more robust and discriminative through additional unsupervised learning. In addition, the proposed method is easy to implement and requires very low extra overhead during the training phase.

In our future studies, we will continue to focus on the use of high-order statistics and investigate other useful multi-task learning strategies for x-vector based speaker verification.

\section{Acknowledgements}

This work was partially funded by the National Natural Science Foundation of China (Grant No. U1836219) and the National Key Research and Development Program of China (Grant No. 2016YFB100 1303).

\bibliographystyle{IEEEtran}

\bibliography{mybib}


\end{document}